\documentclass[mathleft]{an}
\usepackage{times}
\usepackage[dvips]{graphicx}
\usepackage{subfigure}
\usepackage{txfonts}
\usepackage{epsfig}
\usepackage[authoryear]{natbib}

\begin{document}

\title{A modified peak-bagging technique for fitting low-$\ell$ solar p-modes.}

\titlerunning{A modified peak-bagging technique...}

\author{S.T. Fletcher\inst{1}\fnmsep\thanks{Corresponding author:
  \email{s.fletcher@shu.ac.uk}\newline}
\and W.J. Chaplin\inst{2}
\and Y. Elsworth\inst{2}
\and R. New\inst{1}
}

\authorrunning{S. T. Fletcher et. al.}

\institute{Materials Engineering Research Institute, Faculty of
Arts, Computing, Engineering and Science, Sheffield Hallam
University, Howard Street, Sheffield, S1 1WB, UK \and School of
Physics and Astronomy, University of Birmingham, Edgbaston,
Birmingham, B15 2TT, UK }


\abstract{We introduce a modified version of a standard power spectrum
`peak-bagging' technique which is designed to gain some of the
advantages that fitting the entire low-degree p-mode power spectrum
simultaneously would bring, but without the problems involved in
fitting a model incorporating many hundreds of parameters. Employing
Monte-Carlo simulations we show that by using this modified fitting
code it is possible to determine the true background level in the
vicinity of the p-mode peaks. In addition to this we show how small
biases in other mode parameters, which are related to inaccurate
estimates of the true background, are also consequently removed.\\
}

\maketitle

\section{Introduction}\label{SecIntro}

Determining the various parameter values of the resonant modes of
oscillation of the Sun is an important process in the field of
helioseismology. Parameters such as the mode frequencies, lifetimes
and amplitudes can all be used to determine the conditions of the
solar interior. Over the years the quality of helioseismic data has
improved significantly due to the length of data sets increasing,
signal-to-noise ratios being improved and more continuous
observations being made, both from ground-based and space-borne
missions. This has led to the estimated parameter values being
constrained to increasingly greater precision.

As the precision of the parameter estimates increases so more subtle
physical characteristics of the solar interior are being uncovered.
Examples of this include the discovery of asymmetric peaks in the
p-mode power spectrum (e.g., \citealt{Duvall1993};
\citealt{Chaplin1999}), which supported theoretical predictions
(e.g., \citealt{Gabriel1992,Gabriel1995}) and gave evidence to the
belief that acoustic waves are generated within a well-localized
region of the solar interior. Also, long sets of observations have
allowed investigations to be carried out on the dependence of the
mode parameters on solar activity (e.g., for asymmetry dependence
see \citealt{Chaplin2007}).

Determining subtle effects such as these requires analysis
techniques that return robust estimates of the mode parameters. If
this is not the case and the returned parameters are biased, the
inaccuracies can be confused with mode characteristics that have
true physical significance.

Methods of determining solar mode parameters are often referred to
as peak-bagging techniques. For low-degree (low-$\ell$)
Sun-as-a-star observations, a common method of peak-bagging involves
dividing the p-mode power spectrum into a series of `fitting
windows' centered on the $\ell$=0/2 and 1/3 pairs. The modes are
then fitted, pair by pair, to determine how the mode parameters
depend on both frequency (overtone number) and angular degree
without the need to fit the entire spectrum simultaneously.

In this paper we use a Monte-Carlo type approach to estimate the
extent of biases seen in the fitted parameters when using this
`standard' method of fitting, and introduce a new modified technique
designed to reduce these problems.

\section{Fitting Techniques}\label{SecTechniques}

In this section the peak-bagging technique described in the
introduction is explained in more detail. Reasons why the standard
fitting method may return biased parameter estimates are discussed
and the new modified fitting method, designed to limit these
problems, is introduced.

We begin by summarizing briefly the main elements of the standard
peak-bagging method. Within each fitting window the mode peaks are
modeled using a modified Lorentzian equation \citep{Nigam1998}. This
is fitted to the data using an appropriate maximum-likelihood
estimator \citep{Anderson1990}. As we elaborate in
Section~\ref{SecData}, only simulated low-$\ell$ data has been used
in this initial analysis. All modes within the frequency range 1500
$ \leq \nu \leq $ 4600 $\mu$Hz and with angular degree 0 $ \leq \ell
\leq $ 3 were fitted. In \cite{Fletcher2007} it was shown that if
the weak $\ell$ = 4 and 5 modes are not accounted for, they can
often impact on the fitted parameters of the stronger modes. This
was also commented upon in \cite{JimenezReyes2007} and investigated
in \cite{JimenezReyes2008}. Therefore, in the regions of the
spectrum where the modes are strongest, our fitting model was
modified to also include parameters for the $\ell$ = 4 and 5 modes.
This adaption was only made to the model when fitting the $\ell$ =
0/2 pairs, as $\ell$ = 4 and 5 modes never lie within the fitting
windows of $\ell$ = 1/3 pairs.

Within each fitting window the following parameters were varied
iteratively until they converged on their best-fitting values:
\begin{enumerate}
\item A central frequency for each mode. \item A parameter describing
the symmetric rotational splitting pattern for each mode (not
applicable at $\ell$ = 0). \item A linewidth for each mode, for
which the logarithms were varied (the $m$ components in a mode were
assumed to have the same widths). \item A single maximum height --
that of the outer, sectoral $m$ components -- for each mode, for
which the logarithm was varied (the relative $m$ component height
ratios were assumed to take fixed values). \item A single peak
asymmetry parameter (the $m$ components of the modes in the pair
were assumed to have the same asymmetry). \item A flat, background
offset for the fit, whose logarithm was varied.
\end{enumerate}
It should be noted that although the simulated mode peaks were
symmetric in frequency, a parameter characterising asymmetry was
still used so as to be consistent with fits to real data.

For the most part the parameter values returned via this method are
very accurate, especially the mode frequencies. However, there is a
problem associated with restricting the fitting to only a small
slice of the spectrum. For the model to be completely accurate there
would have to be no power within the fitting window from modes lying
outside it, which of course, is not true. The main problem
associated with this is that the background parameter will be
overestimated in order to account for the extra power from the wings
of the outlying modes. The effect has previously been documented in
the literature \citep{Chaplin2003,Fletcher2007} and is illustrated
again in this paper in Section~\ref{SecSimulationFits}.

For fitting purposes the background is usually assumed to be flat in
frequency across the extent of the fitting window. In the case of
the true background this is a fairly safe assumption as the sizes of
the fitting windows tend to be quite small, and over the region of
the spectrum where p modes are observed, the background is thought
to be a relatively weak function of frequency. However, the presence
of the wings from the surrounding modes means the effective
background (i.e., background plus wings) will have a stronger
dependence on frequency. Specifically, the effective background is
observed to rise towards the extreme ends of the fitting window as
these areas are nearer, in frequency, to the surrounding peaks. The
combined effect of incorrectly fitting the true background and using
a model that does not correctly account for the shape of the
effective background, means other parameters may be impacted upon
during the fitting process to minimise these inaccuracies.

\begin{figure*}
\centering{\label{TestFitTop}\includegraphics[width=2.8in]{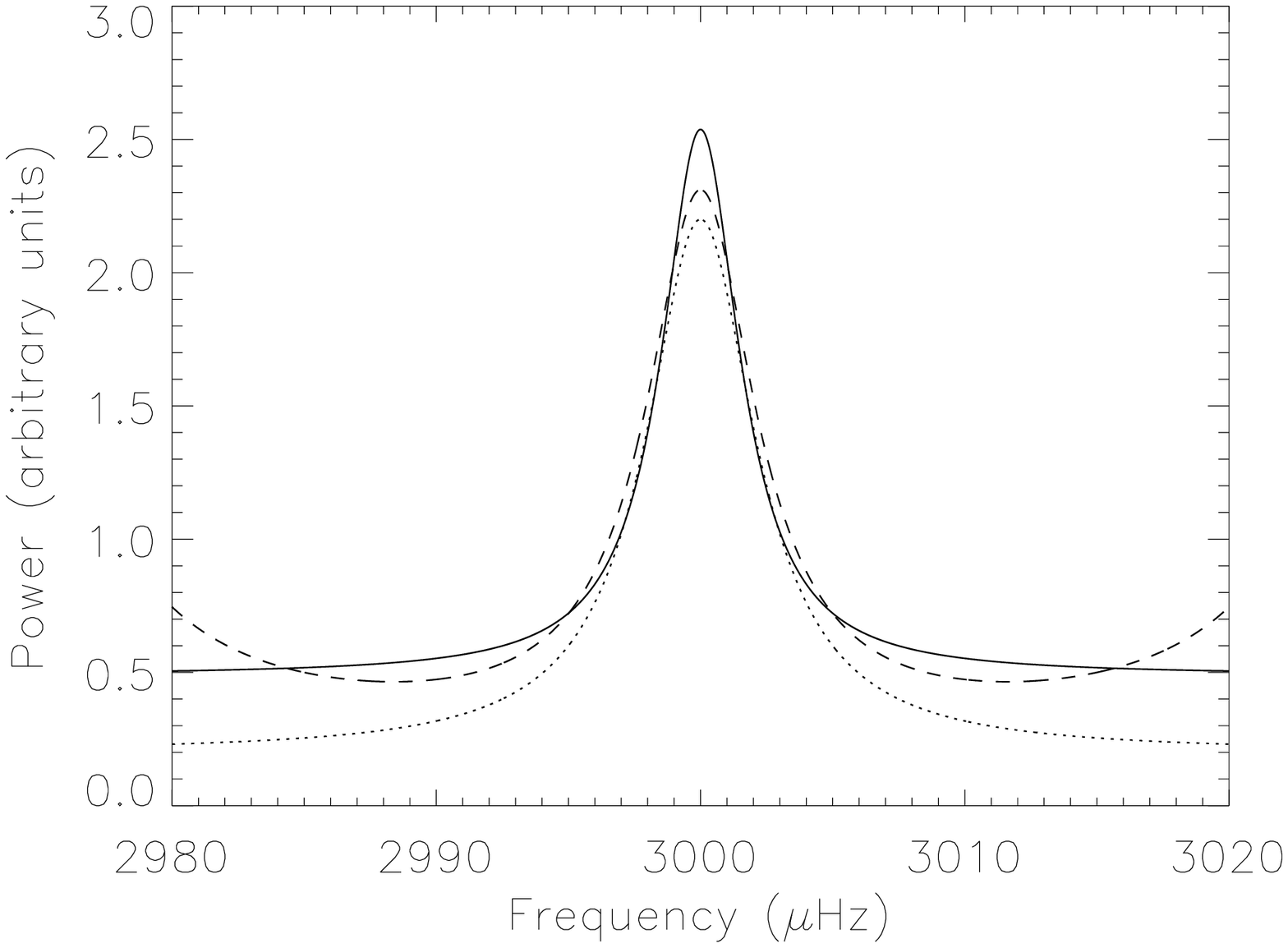}}
{\label{TestFitBot}\includegraphics[width=2.8in]{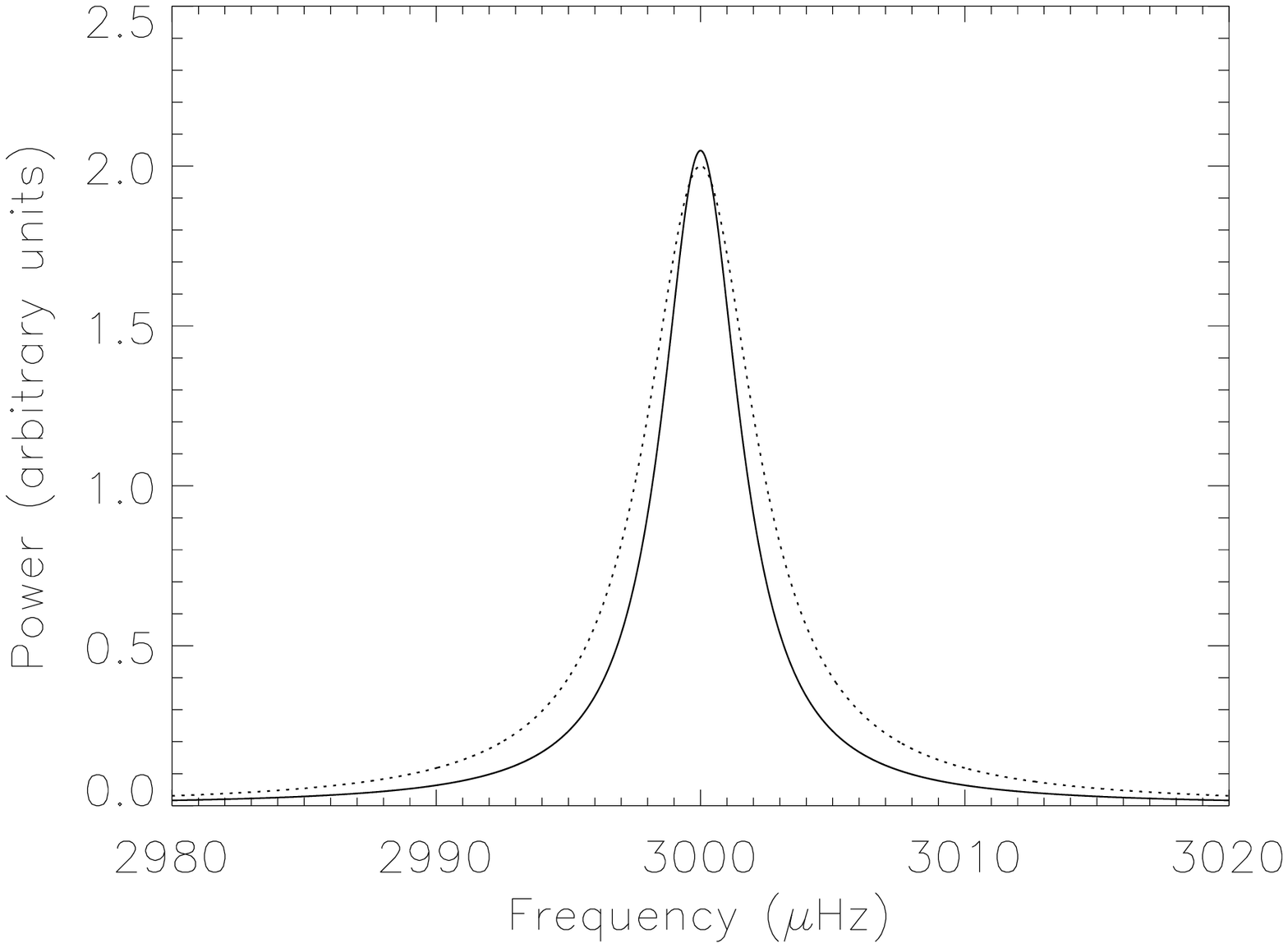}}
\caption{Plot showing fits to a single mode. Large peaks situated
outside the fitting window are not accounted for by the model. In
the left hand plot the dashed line shows the spectrum to be fitted
whereas the solid line gives the fitted result. The dotted line
shows the true peak characteristics without the effects of the
outlying modes. In the right hand plot the solid line shows the
fitted peak minus the fitted background whereas the dashed line
shows the true peak minus the true background. This plot highlights
more clearly the bias in the linewidth and height.} \label{TestFit}
\end{figure*}

This effect is shown in Fig.~\ref{TestFit}, where a simple single
Lorentzian mode peak has been fitted. Either side of this mode,
outside of the fitting window, there are two large peaks the effect
of whose wings can be seen within the fitting window. (This is not a
realistic scenario as the outlying peaks have been substantially
increased in height in order to exaggerate bias in the fits). When a
model is used which does not account directly for the wings of
outlying modes, the fitted background is increased to compensate for
the extra power in order to minimise the likelihood function.
However, because the additional power from the outlying modes is not
flat across the fitting window the likelihood function can be
further minimised by changes in the linewidth and height of the
mode. In this case the background is reduced and the height
increased. This can be seen more clearly in the right-hand plot of
Fig.~\ref{TestFit} where the backgrounds have been removed allowing
the linewidth and height of the fitted peak to be more easily
compared with those of the true peak. We shall see in
section~\ref{SecSimulationFits} that similar biases are seen when
fitting more complicated, more realistic simulated data.

An obvious way of fixing these problems is to fit the entire power
spectrum simultaneously, in which case the wings of all the modes
will be included in the model and the true background will be
fitted. Unfortunately, doing this requires a model with many
hundreds of parameters. As such attempts to fit entire power spectra
would involve large computing times and will be susceptible to
premature convergence. Even so, this technique has been employed
relatively successfully when the time series used were of fairly
short duration (i.e., a few hundred days) (e.g.,
\citealt{RocaCortes1998}; \citealt{Jimenez2002}).

There are at least three possible methods that retain the benefits
of full-spectrum fitting without the need to use a complex model.
One method is to fit the full spectrum, but to describe the
parameters as smooth functions of frequency thus reducing the number
of parameters to be fitted (e.g., \citealt{Jefferies2004}). A second
method is to fit the entire spectrum using a multi-step iterative
process whereby groups of parameters (such as frequencies,
linewidths etc.) are fitted separately. This method has been
pioneered by \cite{Gelly2002}. Finally, a third approach is to
employ an amalgamation of full-spectrum fitting and the standard
`pair-by-pair' fitting. In this scenario one would initially use
parameter fits from the standard technique to create a first-guess
model for the entire spectrum. This model can then be used to fit
just a small slice of the spectrum in the same manner as the
pair-by-pair fitting, allowing only the parameters associated with
the target pair to vary, leaving the remaining model parameters
associated with all the other modes fixed. The advantage of this
approach is that, even though peaks of surrounding modes are not
present in the fitting range, the effects of their wings will still
be modeled.

All of these techniques are worthy of further study, however, in
this paper it is the final of these three approaches that we
investigate. A simple algorithm for this modified fitting technique
is:
\begin{enumerate}
\item Create a model for the full spectrum using the parameters
determined from the standard fitting code. \item Perform a fit
across a single fitting window using the full spectrum model, but
only allow parameters associated with the peaks centered in the
window to vary. \item Repeat the process for all fitting windows.
\end{enumerate}
It should be noted that when forming the full spectrum model from
the parameter estimates determined using the traditional fitting
method, the asymmetry term is reset to zero. This is because the
simplified formalisation of the \cite{Nigam1998} expression used for
our fitting model is only valid within the vicinity of the peaks.
The asymmetry is allowed to vary again during the second round of
fitting.

As in the first set of fits all modes within the frequency range
1500 $ \leq \nu \leq $ 4600 $\mu$Hz and angular degree 0 $ \leq \ell
\leq $ 3 (along with detectable $\ell$ = 4 and 5 modes) were fitted.
However only modes with frequencies up to 4000 $\mu$Hz were analysed
as modes with higher frequencies will begin to experience
contamination from outlying modes that were not fitted using the
pair-by-pair technique. The same parameters that were varied in the
standard fitting techniques were also varied in the modified
version. However, in order to reduce the complexity of the code it
was decided to increase the size of the fitting window to include
both $\ell$ =0/2 and $\ell$ = 1/3 pairs (and consequently the weaker
$\ell$ = 4 and 5 modes as well). Because of the extended size of the
fitting window, it was also decided to allow the background to vary
as a function of one over frequency (which is assumed to be a
reasonable assumption for how the background varies with frequency
in real solar data). This is more important at low frequencies where
the background varies more rapidly with frequency. It should be
noted that this frequency dependence was only applied across the
extent of each fitting window and there were no constraints forcing
a fitting window at higher frequencies to have a smaller background
than was fitted for a window at lower frequencies. However, for the
most part the fitted backgrounds were found to `match-up' from one
window to the next.

\section{Data used}\label{SecData}

A large set of simulated data was created enabling Monte Carlo
simulations to be performed, in order to show whether, and if so
where, the modified fitting code improved upon the standard code.
The data were generated in the time domain using the
first-generation solarFLAG simulator \citep{Chaplin2006}. A database
of mode frequency, power, and linewidth estimates, based on analysis
of spectra made from observations by the Birmingham Solar
Oscillations Network (BiSON), was used in order to fix the various
characteristics of each oscillator component. Visibility levels for
the barely-detectable $\ell$ = 4 and 5 modes were fixed at levels
calculated by \cite{C-Dalsgaard1989}. At the extreme ends of the
spectrum -- where there are no reliable, fitted estimates for the
parameters -- appropriate extrapolations of the known parameter
values were made. In all cases the rotationally induced splitting
between adjacent $m$ components was set at 0.4 $\mu$Hz, again in
order to match fitted estimates from observational data. In all a
sequence of 60 `long' 3456-day time series was generated, along with
240 `short' 796-day time series.

\section{Monte-Carlo simulations}\label{SecSimulationFits}

\begin{figure*}
\centering \subfigure[Background]
{\label{BG}\includegraphics[width=2.2in]{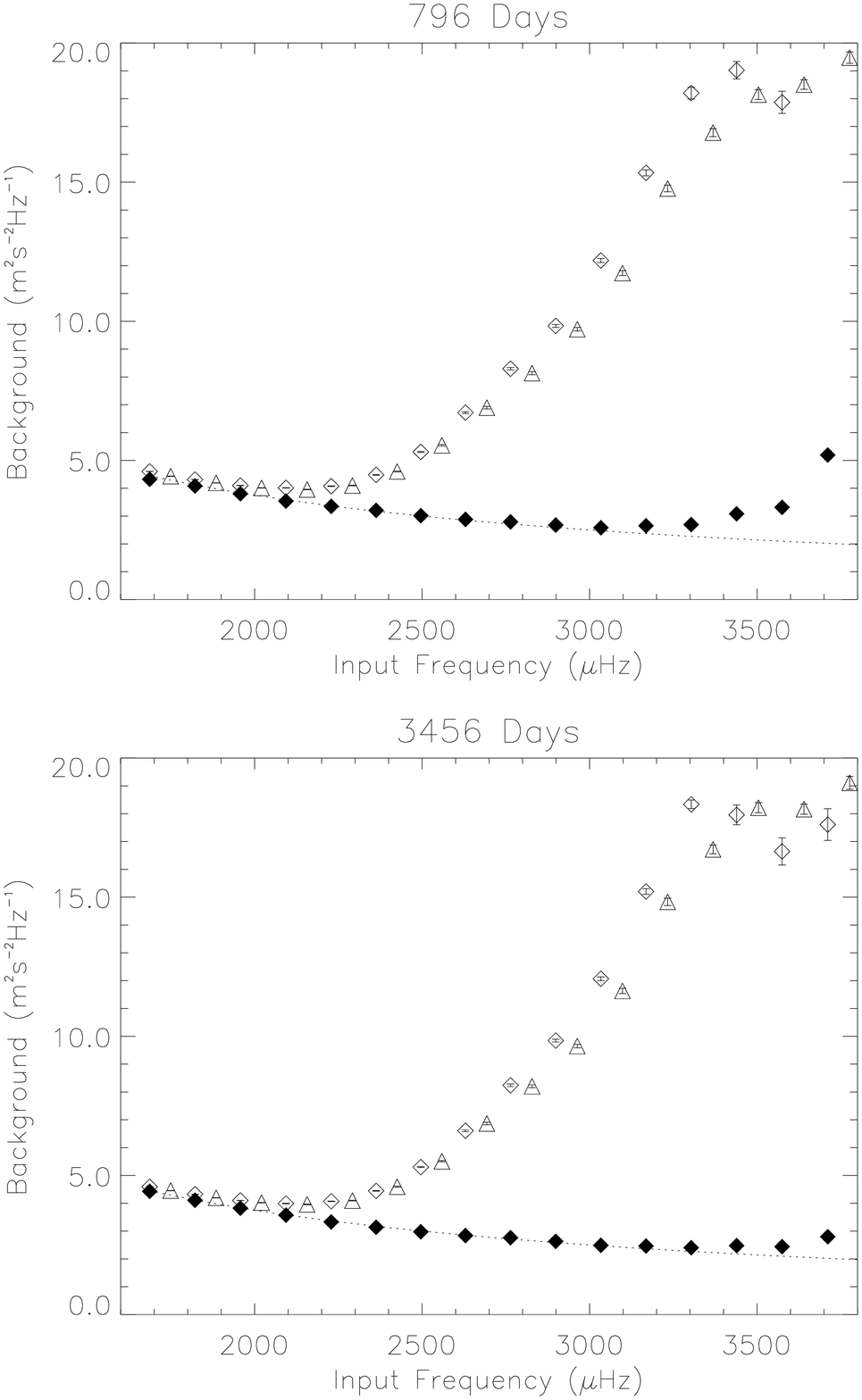}}
\subfigure[Frequency]
{\label{F}\includegraphics[width=2.2in]{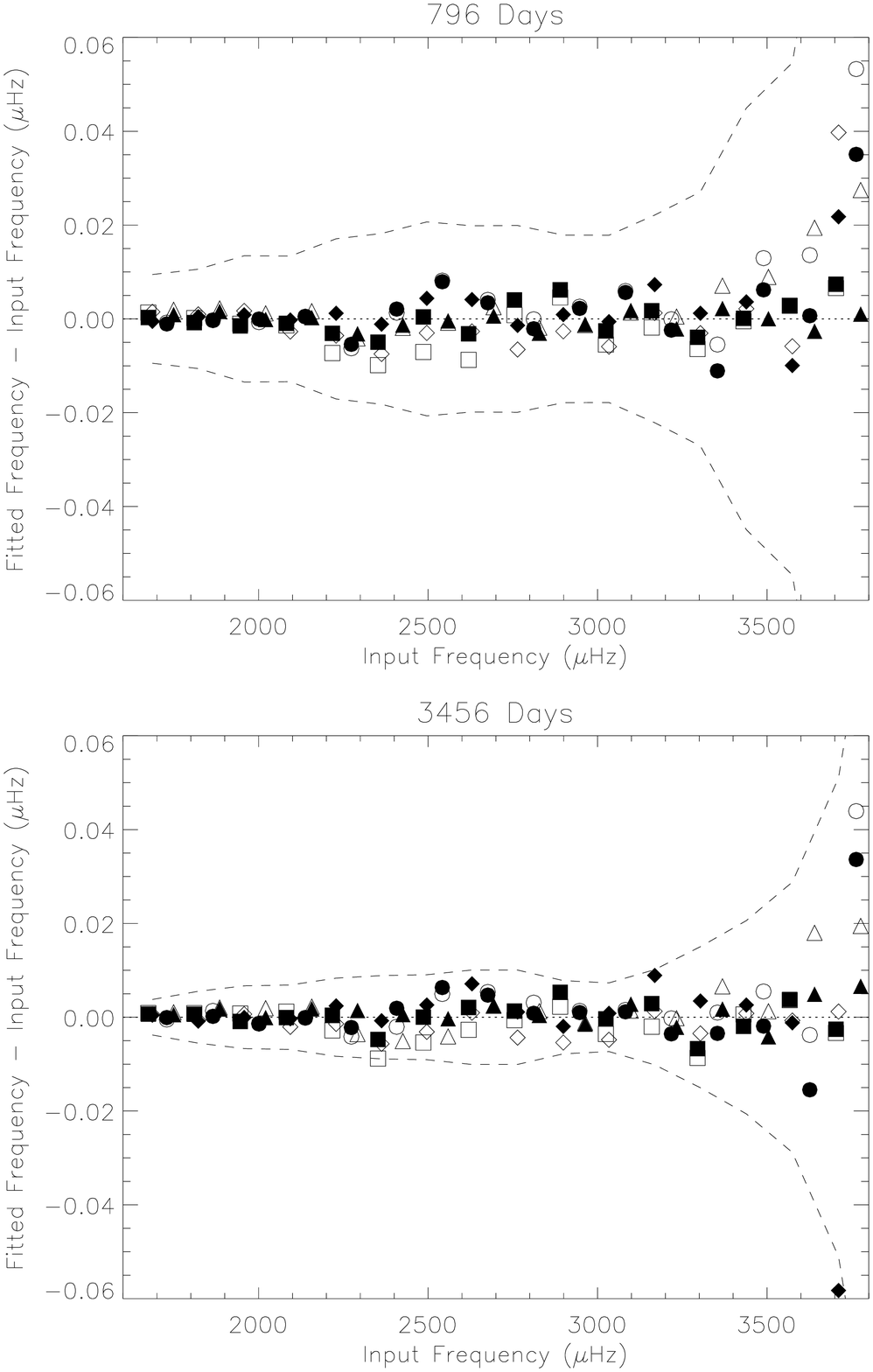}}
\subfigure[Linewidth]
{\label{W}\includegraphics[width=2.2in]{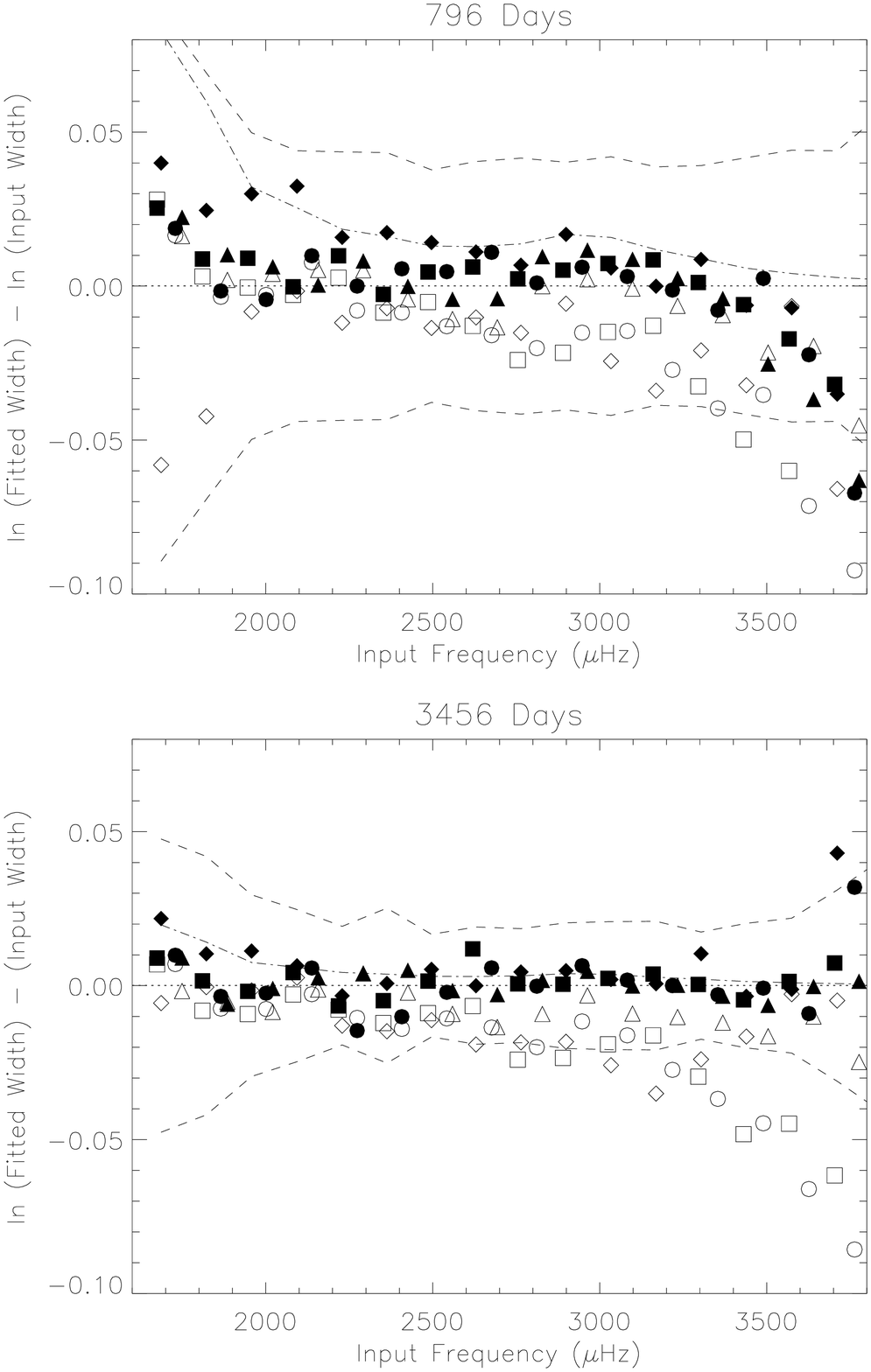}}
\subfigure[Height]
{\label{H}\includegraphics[width=2.2in]{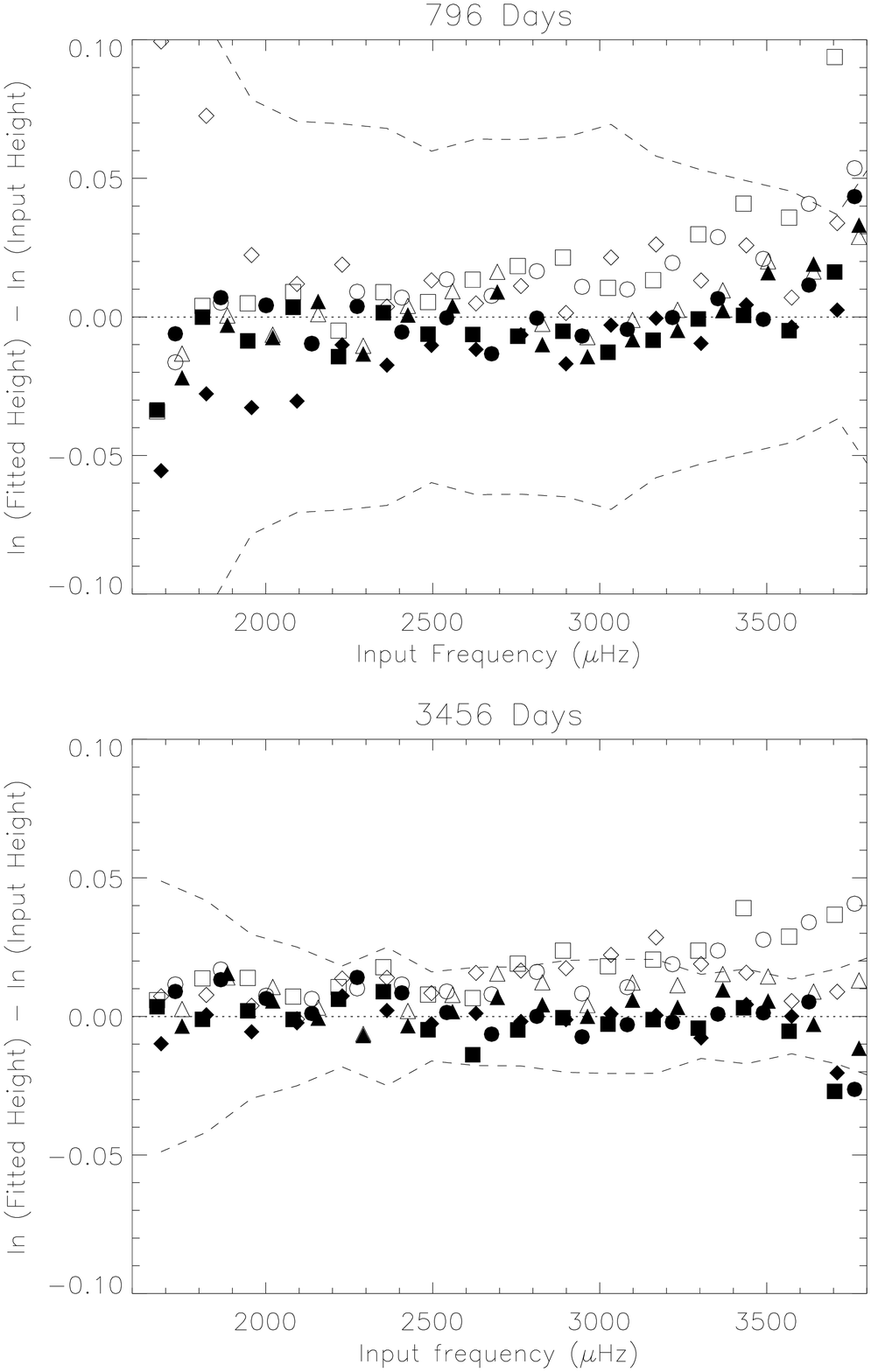}}
\subfigure[Splitting]
{\label{S}\includegraphics[width=2.2in]{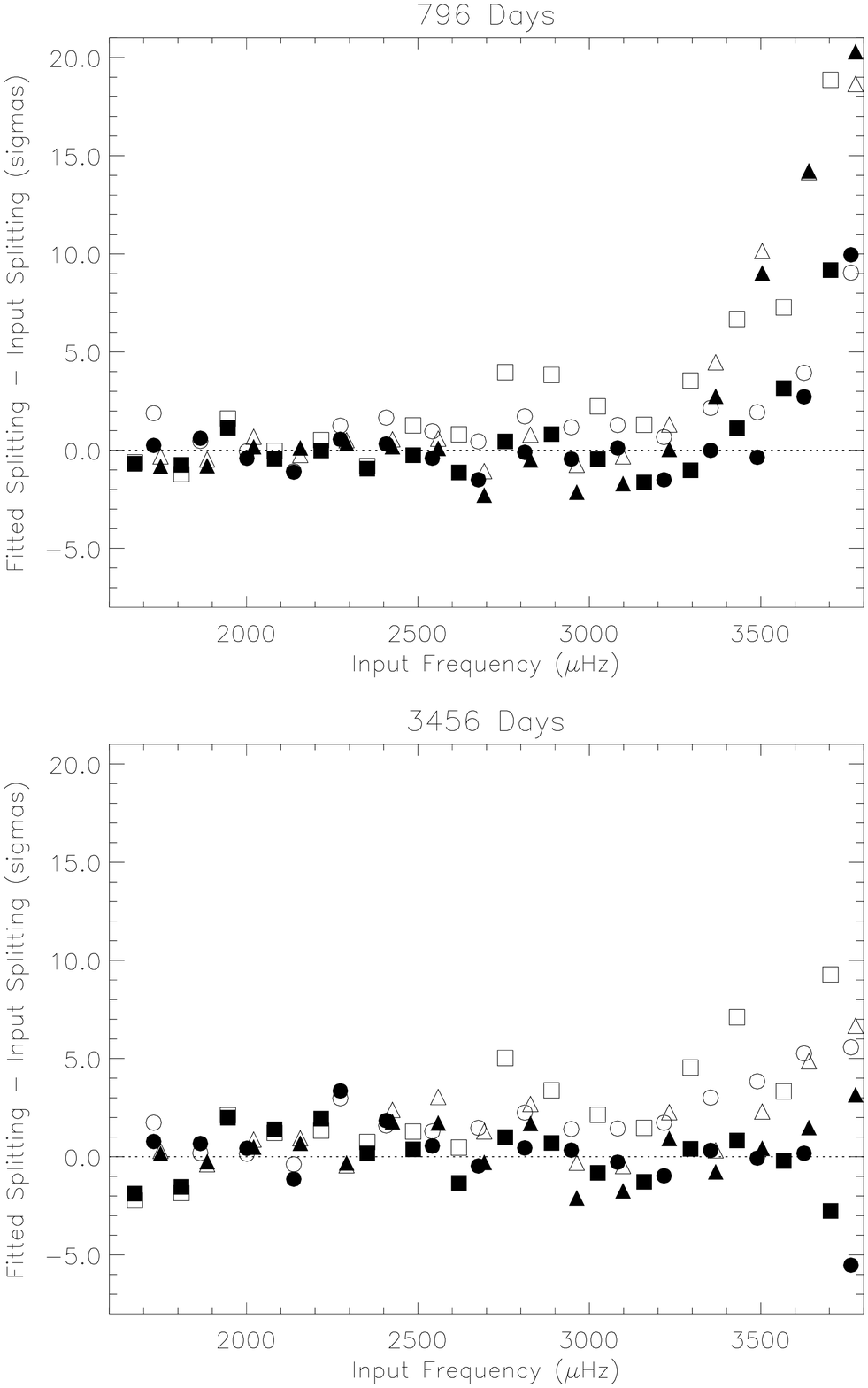}}
\subfigure[Asymmetry]
{\label{A}\includegraphics[width=2.2in]{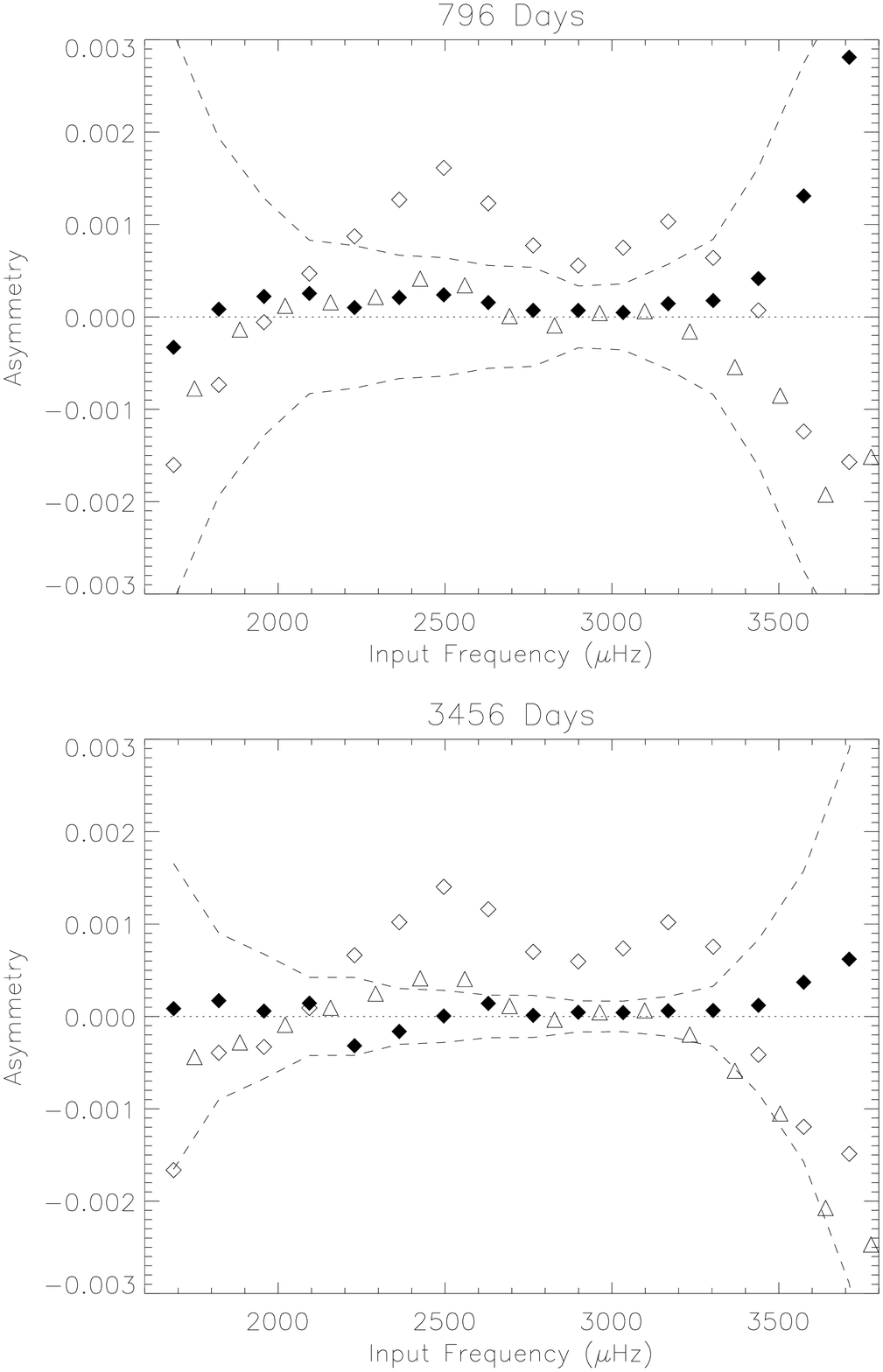}}
\caption{Parameters averaged from fits to 60 sets of simulated
3456-day time series and 240 sets of simulated 796-day time series,
plotted as a function of input frequency. Fitted backgrounds and
asymmetries are plotted directly, whereas frequencies, linewidths,
heights and splittings are plotted in relation to input values (in
the sense fitted - input). Note that for the linewidths and heights
it is the difference in the natural logarithms that are plotted.
Open symbols signify fits returned from the standard fitting code,
whereas solid symbols give results of the modified code. For the
standard code in plots (a) and (f) the fitted parameters for the
$\ell$ = 0/2 window are given by diamonds and by triangles for the
$\ell$ = 1/3 window. In (b) to (e) diamonds signify $\ell$ = 0,
triangles $\ell$ = 1, squares $\ell$ = 2 and circles $\ell$ = 3. The
dashed lines give 0.5-sigma values about the true inputs for a
single fit while the error bars in (a) give 1-sigma errors on the
mean. (0.5-sigma is used as opposed to 1-sigma to give a better
scaling for the plots.) Dotted lines show true input values. In (c)
the dash-dot lines show the expected overestimation of the fits
taking into account the finite resolution of the spectra (see
text).} \label{SimFigs}
\end{figure*}

In Fig.~\ref{SimFigs}, the average fitted parameters determined from
the Monte Carlo testing are shown. We concentrate initially on the
background parameter as this is the parameter for which fits will
most obviously be improved by accounting for the wings of modes that
lie outside the fitting region. In Fig.~\ref{BG}, the average fitted
backgrounds are plotted as a function of frequency for both the
standard and modified fitting methods. For the standard technique
the backgrounds within each $\ell$ = 0/2 and 1/3 fitting window are
plotted, whereas for the modified technique the background levels at
the frequency of each $\ell$ = 0 mode are shown. (Recall that modes
are fitted two pairs at a time by the modified technique and that
the background is allowed to vary as one over frequency across the
fitting window.)

The standard fitting code shows considerable overestimates of the
background, as these estimates also incorporate the wings of the
surrounding modes lying outside the fitting windows. There are also
inconsistences between the fits for different sets of mode pairs,
with the estimated backgrounds in the window centered on $\ell$ =
0/2 pairs being greater on average than those for $\ell$ = 1/3. This
problem can also be attributed to not allowing for the wings of the
surrounding modes, and is a result of two distinct effects. The
first is that the overall power in an $\ell$ = 1/3 pair is slightly
greater than that in an $\ell$ = 0/2 pair (assuming similar mode
frequencies). The second is dependent on the fact that, in the
direction of lower to higher frequencies, $\ell$ = 1/3 mode pairs
are closer to the next 0/2 pair than $\ell$ = 0/2 pairs are to the
next 1/3 pair. This is important because modes at higher frequencies
tend to have larger linewidths and (up until around 3100 $\mu$Hz)
greater heights.

In contrast to the standard fitting, Fig.~\ref{BG} shows that the
modified code returns estimates of the background that closely match
the input values, at least up to frequencies of around 3000 $\mu$Hz.
The small overestimate of the background returned above this
frequency is not entirely understood, although it does appear from
the plots that the effect is dependent on the length of the time
series.

It has previously been documented that many of the fitted mode
parameters are correlated with one another (e.g.,
\citealt{Fletcher2007}). This suggests that the effect of not
accurately fitting the true background, when using the standard
fitting technique, may also impact on the estimates of the other
parameters. This was highlighted in Section~\ref{SecTechniques} in
Fig~\ref{TestFit}, where the height and widths were shown to be
biased due to the presence of unmodeled outlying peaks.

In Fig.~\ref{F} average differences between the fitted and input
frequencies are shown. The plot shows that for the most part the
frequencies returned by the standard fitting technique are both
robust and accurate and so there is little to improve upon when
using the modified fitting code. The frequencies show the smallest
correlation with the other parameters and therefore one might expect
them to be particularly robust when fitting the data using the
standard method.

In Fig.~\ref{W} the average differences in the natural logarithms of
the fitted and input linewidths are shown. This time there is a
small but clear systematic bias in the estimates returned by the
standard fitting method. As the magnitude of the bias is quite small
the differences in the natural logarithms closely approximate the
fractional bias in the linear width values, (i.e., a difference in
the natural logarithm of -0.05 indicates the fitted linear values of
the widths underestimate the input value by about 5 percent). For
the 3456-day power spectra, this bias is clearly reduced when using
the modified fitting code, with the largest improvements coming at
higher frequencies and for the higher-degree ($\ell$ = 2 and 3)
modes.

However, the results of the fits to the 796-day power spectra are
not quite as clear cut. Over a large part of the frequency range the
standard fitting code still returns fits with a negative bias, and
for the most part (between around 2500 and 3500 $\mu$Hz) the
modified fitting code improves upon this. However, there is evidence
of a small positive bias in results from the modified code which
increases at lower frequencies and for lower-degree modes. This is
believed to be a resolution issue. It has previously been shown that
when fitting the linewidth of the modes in the power spectrum, the
estimates returned are actually better matches to the true inputs
plus the binwidth (see \citealt{Chaplin1997}). Since the binwidth is
larger for a power spectrum made from shorter times series, this
effect is more easily observed in the 796-day data. The
non-horizontal dotted lines in Fig.~\ref{W} give, for the $\ell$ = 0
modes, the difference between the natural logarithm of the linewidth
plus the binwidth and the natural logarithm of the linewidth only.
As such, this gives the line we would expect the $\ell$ = 0 points
to fall upon (assuming all effects other than this resolution bias
have been removed) and the plot shows that over much of the
frequency range this is indeed the case for the modified fitting
code. The overestimation is reduced by the square root of the number
of peaks being fitted which explains why the effect is reduced for
the higher-degree modes.

When fitting the power spectrum, one will nearly always find a very
strong correlation between the estimated linewidths and heights of
the modes. Therefore, much the same pattern as was seen for the
linewidths will be seen for the heights but with the opposite bias.
This is indeed seen to be the case as shown in Fig.~\ref{H}. Again
the modified code is seen to reduce the overall extent of the bias
especially with the longer 3456-day times series.

The results of fitting the rotational splittings are shown in
Fig.~\ref{S}. In this plot the differences between averages of the
fitted splittings and the true input values are divided by the error
on the mean. This has been done to more easily display the
splittings at all frequencies and angular degrees on the same scale.
It has been well documented that at high frequencies the fitted
splittings will tend to overestimate the true values (eg.,
\citealt{Chaplin2001}). This is a result of the strong mode blending
that occurs as the linewidths of the modes increase at higher
frequencies. This problem tends to be larger for lower-degree modes
as the separation in frequency between the outermost components is
smaller. However, this effect is less obvious when plotting the
differences and dividing by the error, since the uncertainties on
the lower-degree splittings are also very large.

Fig.~\ref{S} clearly shows a distinct overestimate of the fitted
splittings at high frequencies with a number of points showing
values larger than 3 sigma. As with the previous parameters, over
the medium to high-frequency ranges, there appears to be a
significant reduction in the bias when using the modified fitting
code. However, the extremely large overestimates that are seen when
fitting the highest frequencies are not significantly reduced.

The final parameter investigated was the peak asymmetry. Even though
no asymmetry was included in the simulations, it is still important
to investigate in order to check that the model returns estimates
consistent with zero. Fig.~\ref{A} shows the average fitted
asymmetries as a function of frequency. Unlike the previous
parameters the asymmetry was kept constant throughout the fitting
window, meaning there is no $\ell$-dependence and thus fewer points
in the plot. While the intrinsic values of the average fitted
asymmetries are actually very small, they are significant and hence
not consistent with zero. For the standard fitting code there are
two sets of values for the asymmetries, one for the $\ell$ = 0/2
pairs and one for the 1/3 pairs. This is the same scenario as for
the backgrounds, and as for that parameter, there is a significant
disparity between the average fitted asymmetries for the two
different sets of mode pairs, with the $\ell$ = 0/2 generally
showing a much larger bias than the $\ell$ = 1/3. In fact the
disparities seen in the background and asymmetry are most likely
related.

The modified code seems to reduce the overall bias in the fits,
although in this case, it is not eliminated entirely. Also, the
plots show that the extent of the bias has a fairly smooth response
as a function of frequency and this trend is similar for all cases
(although somewhat exaggerated for the fits to the $\ell$ = 0/2
pairs). These two facts would suggest that there is some underlying
cause for the bias in the fitted asymmetries that is not being
correctly addressed, even with the modified fitting technique.

\section{Conclusions}\label{SecConclusion}

We have introduced a new peak-bagging fitting technique that was
designed in order to gain some of the advantages that fitting the
entire `Sun-as-a-star' p-mode spectrum simultaneously might bring.
Results on Monte-Carlo simulations demonstrated that this modified
fitting method enabled accurate estimates of the true background
level in the vicinity of the p-modes to be determined, something
that was not possible using previous peak-bagging techniques. The
Monte-Carlo simulations also showed that small biases present in
other parameters (e.g, heights and widths) can be reduced using the
modified code.

If the modified fitting code is indeed giving the true background
level then it will enable investigations of the noise
characteristics to be carried out within the frequency range of the
p modes. Previously, study of noise characteristics (which include
both solar and instrumental sources as well as atmospheric sources
in the case of ground based instruments) was limited to measurements
of the background at frequencies well above and below the main part
of the p-mode spectrum (e.g., \citealt{Chaplin2005}).

The modified code has also been tested using data collected by the
Global Oscillations at Low Frequencies (GOLF) instrument on board
the Solar and Heliospheric Observatory (SOHO) spacecraft. The
results of this analysis will be presented in a subsequent paper.
Additionally the modified code has been adapted for use with time
series that contain breaks, such as those collected by the ground
based BiSON group. Analysis of these type of data (both simulated
and real) is the subject of current work.

\section*{Acknowledgments}

STF acknowledges the support of the School of Physics and Astronomy
at the University of Birmingham. We thank all those associated with
BiSON which is funded by the Science and Technology Facilities
Council (STFC). The authors are grateful for the support of the
STFC. We acknowledge the Solar Fitting at Low Angular degree Group
(solarFLAG) for use of their artificial helioseismic data.

\end{document}